\documentclass[aps,prb,reprint,showpacs,superscriptaddress,amsmath]{revtex4-1}
\usepackage{color}
\usepackage{graphicx}
\begin{document}

\title{Variation of transition temperatures and residual resistivity ratio in vapor-grown FeSe}

\author{A.~E.~B\"ohmer }
\email[corresponding author: ]{aboehmer@iastate.edu}
\affiliation{Ames Laboratory US DOE, Iowa State University, Ames, Iowa 50011, USA}

\author{V.~Taufour }
\affiliation{Ames Laboratory US DOE, Iowa State University, Ames, Iowa 50011, USA}

\author{W. E. Straszheim}
\affiliation{Ames Laboratory US DOE, Iowa State University, Ames, Iowa 50011, USA}

\author{T. Wolf }
\affiliation{Karlsruhe Institute of Technology, Institut f\"ur Festk\"orperphysik, 76021 Karlsruhe, Germany}

\author{P.~C.~Canfield}
\affiliation{Ames Laboratory US DOE, Iowa State University, Ames, Iowa 50011, USA}
\affiliation{Department of Physics and Astronomy, Iowa State University, Ames, Iowa 50011, USA }

\date{\today}

\begin{abstract}
The study of the iron-based superconductor FeSe has blossomed with the availability of high-quality single crystals, obtained through flux/vapor-transport growth techniques below the structural transformation temperature of its tetragonal phase, $T\approx 450^\circ$C. Here, we report on the variation of sample morphology and properties due to small modifications in the growth conditions. A considerable variation of the superconducting transition temperature $T_{\mathrm c}$, from 8.8 K to 3 K, which cannot be correlated with the sample composition, is observed. Instead, we point out a clear correlation between $T_{\mathrm c}$ and disorder, as measured by the residual resistivity ratio. Notably, the tetragonal-to-orthorhombic structural transition is also found to be quite strongly disorder dependent ($T_{\mathrm s}\approx 72-90$ K), and linearly correlated with $T_{\mathrm c}$.  
\end{abstract}


\maketitle

\section{Introduction}

Among the Fe-based superconductors, the binary FeSe has the simplest crystallographic structure and some of the most intriguing properties. FeSe in its tetragonal, PbO-type structure, is superconducting with a $T_{\mathrm c}\approx8$ K at ambient pressure\cite{Hsu2008}, which is enhanced four-fold up to 37 K under pressure\cite{Medvedev2009}. Furthermore, FeSe exhibits an intensively studied nematic (i.e., orthorhombic and paramagnetic) phase that, unusually, extends from a tetragonal-to-orthorhombic structural transition at $T_{\mathrm s}\approx90$ K down to $T_{\mathrm c}$ at ambient pressure\cite{McQueen2009,Watson2015}. FeSe also features a rather elusive, small-moment magnetically ordered phase\cite{Bendele2010,Bendele2012,Sun2015,Kothapalli2016} induced by pressure. Extremely small Fermi surfaces, related to strong orbital selective electronic correlations\cite{Maletz2014}, place superconductivity in FeSe in the vicinity of the interesting BCS-BEC crossover regime\cite{Kasahara2014}. The detailed study of all these properties was facilitated, and in some cases only made possible, by the availability of high-quality single crystals. Notably, single-crystal preparation is complicated by the rather complex binary Fe-Se composition-temperature phase diagram\cite{Okamoto1991}. In particular, the superconducting tetragonal PbO-type phase of FeSe has only a very narrow range of stability and undergoes a phase transformation on warming above 457$^\circ$C. In consequence, any preparation procedure above this temperature yields samples not formed in the tetragonal phase, that structurally transform upon cooling to room temperature. This inevitably leads to impurity phases and internal strains, thus reducing crystal quality. 

Early studies of FeSe used polycrystalline material prepared by solid state synthesis\cite{Hsu2008,Mizuguchi2008,Margadonna2008,Pomjakushina2009,McQueen2009II}. In particular, the detailed investigation in Ref.  \onlinecite{McQueen2009II} shows how the properties of these polycrystalline samples are affected by annealing at temperatures between 300$^\circ$C-450$^\circ$C. There are also several early studies of the growth of tetragonal FeSe using Cl-salt-based flux techniques\cite{Zhang2008,Braithwaite2009,Hu2011} and chemical vapor transport\cite{Patel2009,Hara2010}. In many of these studies, it was observed that crystals (which formed at $T>450^\circ$C) have a (partially) hexagonal habit and are composed of both hexagonal and tetragonal phases, a consequence of the phase transformation described above. AlCl$_3$ has been known for many years as a transport agent for metals, selenides and sulfides \cite{Moh1971,Lutz1974}. A breakthrough came with the use of a eutectic mix of KCl and AlCl$_3$ salts with low melting temperature to obtain FeSe directly in its tetragonal phase in flux\cite{Chareev2013} or vapor transport\cite{Boehmer2013} techniques below 450$^\circ$C. This significantly improved the crystal quality, as shown by an approximately ten-fold increase in residual resistivity ratio with respect to the previously available samples\cite{Kasahara2014}. 
In this report, we describe how changes in the conditions of the sample growth influence the morphology and the properties of the obtained material and point out a correlation between residual resistivity ratio, $T_{\mathrm s}$, and $T_{\mathrm c}$.

 
\section{Experimental details}

Single crystals of FeSe were prepared by a chemical vapor transport technique using elemental Fe and Se and a eutectic mix of the chlorine salts, KCl and AlCl$_3$ (molar ratio 1:2), in a constant temperature gradient. The furnace was tilted at an angle of $\sim15^\circ - 20^\circ$ to enhance convection (see Fig. 1). The Fe:Se ratio, the temperature conditions and the amount of starting materials were varied. A total of more than 20 batches were studied. Powder x-ray diffraction was performed using a Rigaku Miniflex II diffractometer with Cu K$_\alpha$ radiation. The compositions of crystals from three batches were refined using wave-length dispersive x-ray spectroscopy (WDS). Magnetization was measured using a Quantum Design MPMS SQUID in a field of 20 Oe under zero-field cooled conditions and with arbitrary orientation of the often very small single crystals. Multiple pieces were checked for each batch. Electrical resistance was measured using a LR-700 resistance bridge and contacts were made with silver epoxy and silver paint. 

\section{Results of growth experiments}

\begin{figure}
	\includegraphics[width=8.6cm]{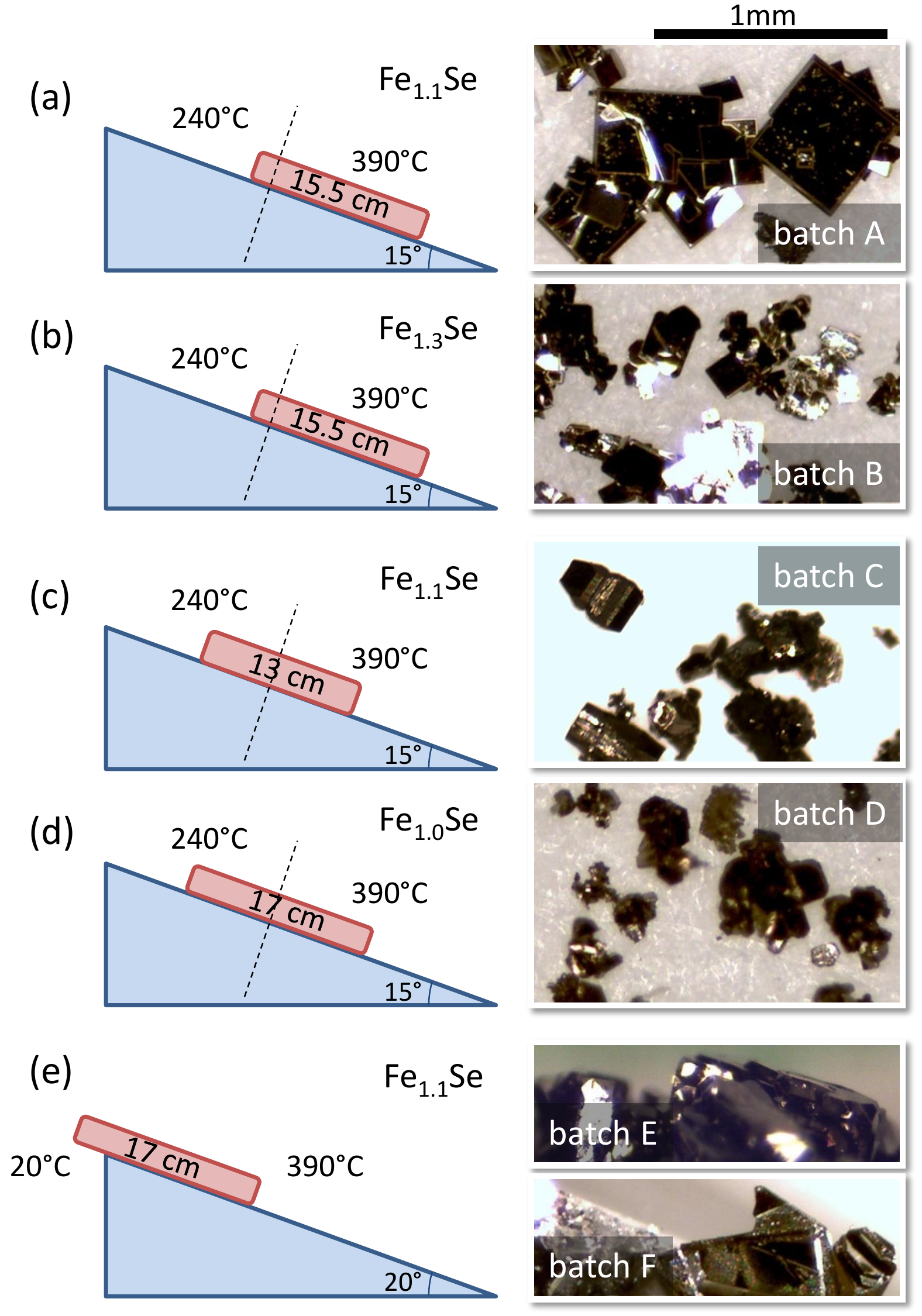}
	\caption{Schematics and photographs of the growth for several batches of FeSe. Batches A-D in panels (a)-(d) were prepared in a tilted two-zone furnace in a silica ampoule of 14 mm inner diameter, using only 20 mg of Fe powder as starting material, diluted in 1 g of a mix of KCl and AlCl$_3$. Fe:Se ratio, ampoule position and ampoule length were varied. The dashed line represents the boundary of the furnace's two zones. The two batches E and F in panel (e) were prepared with 250 mg of Fe in a tilted tube furnace using a larger and less well controlled temperature gradient that was created by letting one end of the ampoule stick out into room temperature conditions. For batch F, lump iron instead of Fe powder was used.}
	\label{fig:1}
\end{figure}

A series of growths with a small amount of starting materials (20 mg Fe powder) diluted in 1 g of KCl/AlCl$_3$ were studied. The study of this series was instigated by the need to optimize and reproducibly grow crystals of isotopically pure $^{57}$FeSe for synchrotron M\"ossbauer spectroscopy\cite{Kothapalli2016}. Such growths had to be performed in a small batches due to the cost of $^{57}$Fe. The starting materials were sealed in a silica ampoule of 13-17 cm length and 14 mm inner diameter. The ampoules were placed in a two-zone furnace with two heater coils of 10 cm length each located 4 cm apart from each other. The growth time was typically 10-14 days. Results from a few selected experiments are presented in Fig. 1 (a)-(d). As shown in Fig. 1 (a), we successfully obtained tetragonal plate-like single crystals using this small-scale experiment with a ratio of Fe to Se powder of 1.1:1 and with the two zone temperatures set to 240$^\circ$C and 390$^\circ$C (batch A). To test the sensitivity to the starting composition, batch B was prepared with an Fe:Se ratio of 1.3:1 (Fig. 1 (b)). Still, well-defined platelet-like single crystals, albeit smaller than for batch A, were obtained. In contrast, by simply changing the position of the ampoule in the furnace as indicated in Fig. 1 (c), crystals of different morphology, namely rod-like with occasionally well-formed side facets, were obtained. This shift of the ampoule further into the 240$^\circ$C zone of the furnace most likely brings the actual temperature of the cold end closer to 240$^\circ$C. Powder x-ray diffraction confirms this batch C also to be in the PbO-type tetragonal phase. Finally, Fig. 1 (d) shows an experiment with a lower Fe:Se ratio of 1:1. A mixed phase batch is obtained, which contains samples of hexagonal and of tetragonal morphology. The hexagonal samples are attracted to a magnet at room temperature. As shown by the compositional analysis below, these are likely composed of the known ferrimagnetic Fe$_7$Se$_8$ phase\cite{Okazaki1961}. These experiments show that a small Fe excess is important to suppress the formation of the hexagonal phase, even though the amount of excess Fe is less critical. The sensitivity to the ampoule position in the furnace further indicates that the growth is very sensitive to temperature conditions.

\begin{figure}
	\includegraphics[width=8.6cm]{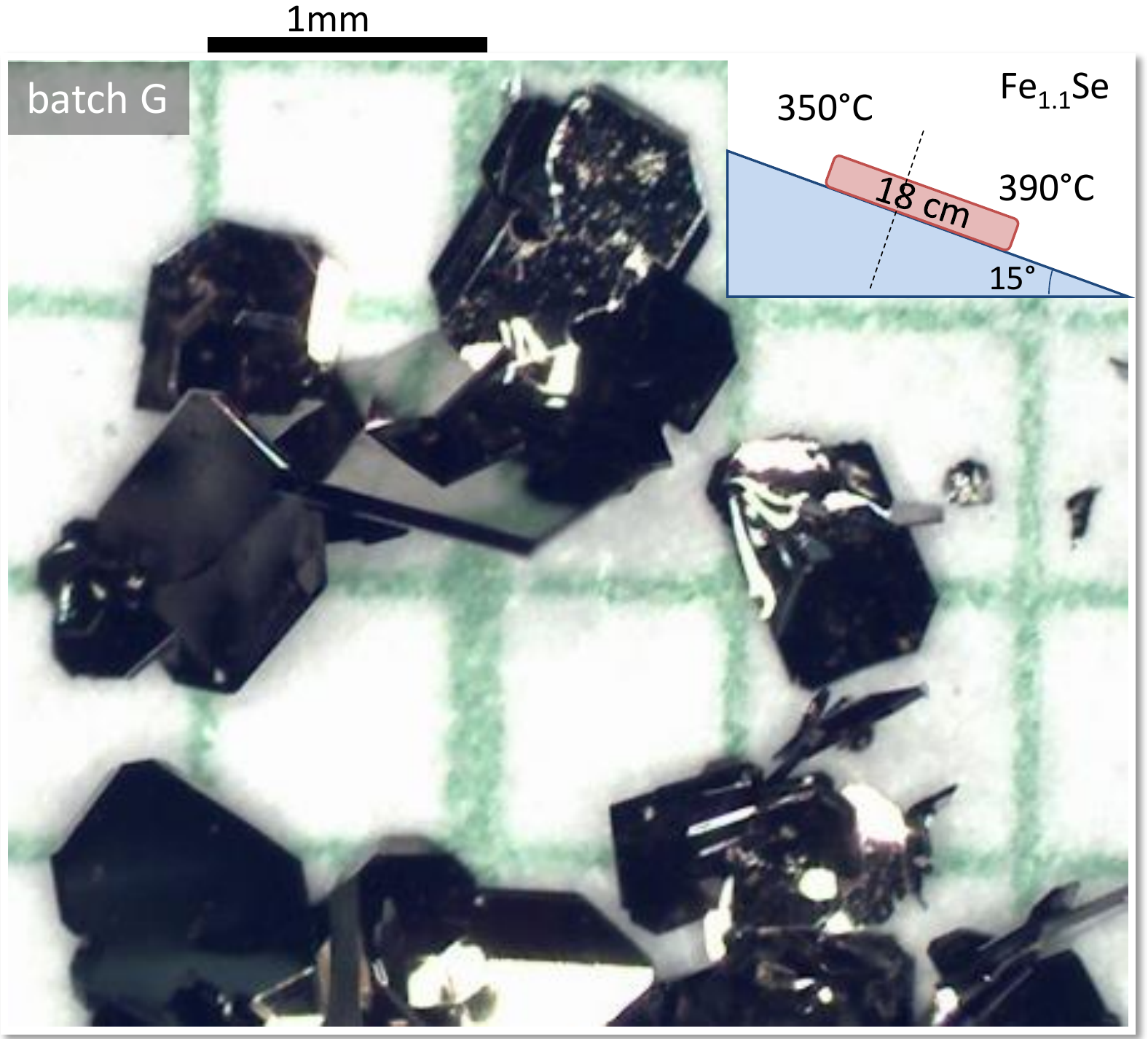}
	\caption{Photograph of representative samples from the reference batch G containing mm-sized platelet-shaped single crystals of tetragonal FeSe. The batch was prepared in a two-zone furnace with a smaller, constant temperature gradient of 350$^\circ$C to 390$^\circ$C over two weeks. As starting materials, Fe and Se powder (total mass $\sim0.5$ g) in an molar ratio of 1.1:1 were diluted in $\sim5$ g of a eutectic mix of KCl and AlCl$_3$.}
	\label{fig:2}
\end{figure}

\begin{figure}
	\includegraphics[width=8.6cm]{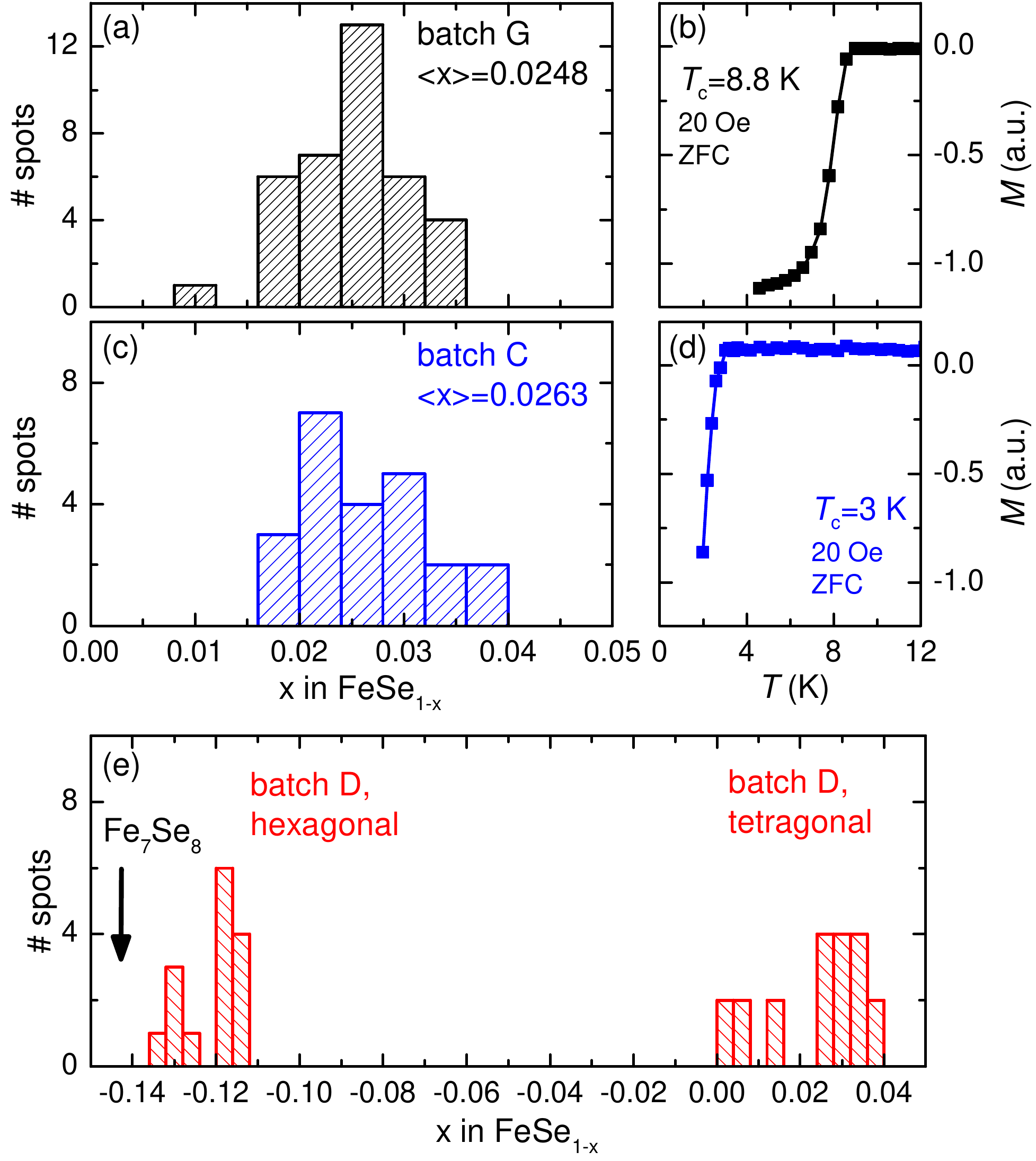}
	\caption{Histogram of the WDS results from multiple spots on FeSe single crystals of (a) reference batch G, (c) batch C and (e) batch D. 4-5 freshly cleaved samples per batch were measured, evaluating only flat homogeneous regions. The average value of the Se deficiency $<x>$ in FeSe$_{1-x}$ is indicated. (b), (d) Representative zero-field cooled magnetization curves indicating $T_{\mathrm c}=8.8$ K for batch G and $T_{\mathrm c}=3$ K for batch C.}
	\label{fig:3}
\end{figure}

 Fig 1 (e) shows batches E and F which were prepared using even more extreme, and far less controlled, temperature gradients in a 36 cm long tilted tube furnace with one end of the ampoules sticking out of the furnace and having temperatures of $\sim150^\circ$C-200$^\circ$C. For batch F, additionally, lump iron instead of Fe powder was used. In both cases, the amount of starting materials was larger with 250 mg of Fe, mixed with Se in a molar ratio of 1.1:1, and 5 g of KCl/AlCl$_3$ mix. As shown below, these growths are examples of the largest variation of crystal quality. 
 
Having observed that the size of the temperature gradient has a large effect on size and morphology of the single crystals grown, we prepared another batch (batch G, shown in Fig. 2) using a smaller and well-controlled temperature gradient, namely 350$^\circ$C and 390$^\circ$C in the tilted two-zone furnace. The amount of starting materials was 250 mg of Fe, mixed with Se in a molar ratio of 1.1:1 and 5 g of KCl/AlCl$_3$. The single crystals from this batch have homogeneous properties and a morphology similar to the high-quality single crystals reported elsewhere (e.g., Refs.  \onlinecite{Kasahara2014,Watson2015,Sun2015}). They  were used for studies in Refs. \onlinecite{Tanatar2015,Kaluarachchi2016,Kothapalli2016,Teknowijoyo2016} and have a structural transition at $T_{\mathrm s}=87-89$ K, a superconducting transition temperature of $T_{\mathrm c}=8.7-8.8$ K and a ratio between the resistance at 300 K and the resistance just above $T_{\mathrm c}$ of $\sim 25$. Here, batch G will serve as a reference for composition and physical properties. 
We note that samples in Refs. \onlinecite{Boehmer2013,Boehmerthesis} were prepared in a temperature gradient created by placing the ampoule with one end close to the opening of a single-zone furnace. Below, we compare the transition temperatures of these earlier samples with those of the batches G,E and F.

It has been reported in a study of polycrystalline samples that $T_{\mathrm c}$ varies sensitively as a function of sample composition, $x$ in FeSe$_{1-x}$ (Ref. \onlinecite{McQueen2009II}). In particular, it was shown that $T_{\mathrm c}$ is highest for samples with the lowest degree of off-stoichiometry and the changes in composition have been related to the synthesis/annealing temperature.  
In Fig 3. (a)-(d), we show composition and superconducting transition of two contrasting batches, batch C and the reference batch G. We find that the $T_{\mathrm c}$ of batch G is reproducibly $8.7-8.8$ K, however, samples from batch C have reproducibly a very low $T_{\mathrm c}<5$ K. Using WDS, the composition of 4-5 samples of each batch (2-15 clean, flat spots on each sample) was determined and the histogram of the results is shown in Fig. 3 (a), (c). In contrast to the expectation, WDS yields the same composition for the two batches within error, namely Fe:Se=1:0.975(6). We note that the composition as determined by WDS may have a larger systematic error, which does not, however, affect the comparison between the two batches. For comparison, a different technique, namely full structural refinement of single-crystal four-circle x-ray data in Refs. \onlinecite{Boehmer2013,Boehmerthesis} yielded compositions closer to stoichiometry, Fe:Se = 0.995(5):1, for three vapor-grown single crystals with $T_{\mathrm s}=87-90$ K and $T_{\mathrm c}=7.75-9$ K.  

The composition histogram for batch D, which contains samples both of tetragonal and hexagonal morphology, is shown in Fig. 3 (e). The two phases are also clearly distinguished from their composition. Samples of hexagonal morphology have a significant Se excess and are in composition close to the reported, room-temperature ferrimagnetic Fe$_7$Se$_8$ phase\cite{Okazaki1961}, whereas samples with tetragonal morphology are similar in composition to the other batches. The superconducting transition temperature of the tetragonal samples of this batch varies considerably between $3.8-8.6$ K. 

\section{Resistance: Results and discussion}

Being unable to relate the large $T_{\mathrm c}$ variation to compositional changes within the of resolution of our measurements, we consider whether disorder might be the dominant factor in determining $T_{\mathrm c}$. It is well-known that $T_{\mathrm c}$ is very sensitive to disorder in many unconventional superconductors. A simple measure for disorder is the residual resistivity ratio (RRR). Below, we use the ratio between the electrical resistance at $T=250$ K and the resistance just above $T_{\mathrm c}$ for the RRR value. Since batches like C, prepared with only 20 mg of Fe powder, do not contain samples large enough for resistance measurements, we turn to two other batches (E and F, showing in Fig. 2 (e)), which were intentionally prepared as to contain 'lower-quality' single crystals by using a large temperature gradient. For batch F, we additionally used lump Fe instead of Fe powder, which resulted in samples of significantly varying $T_{\mathrm c}$.

\begin{figure}
	\includegraphics[width=8.6cm]{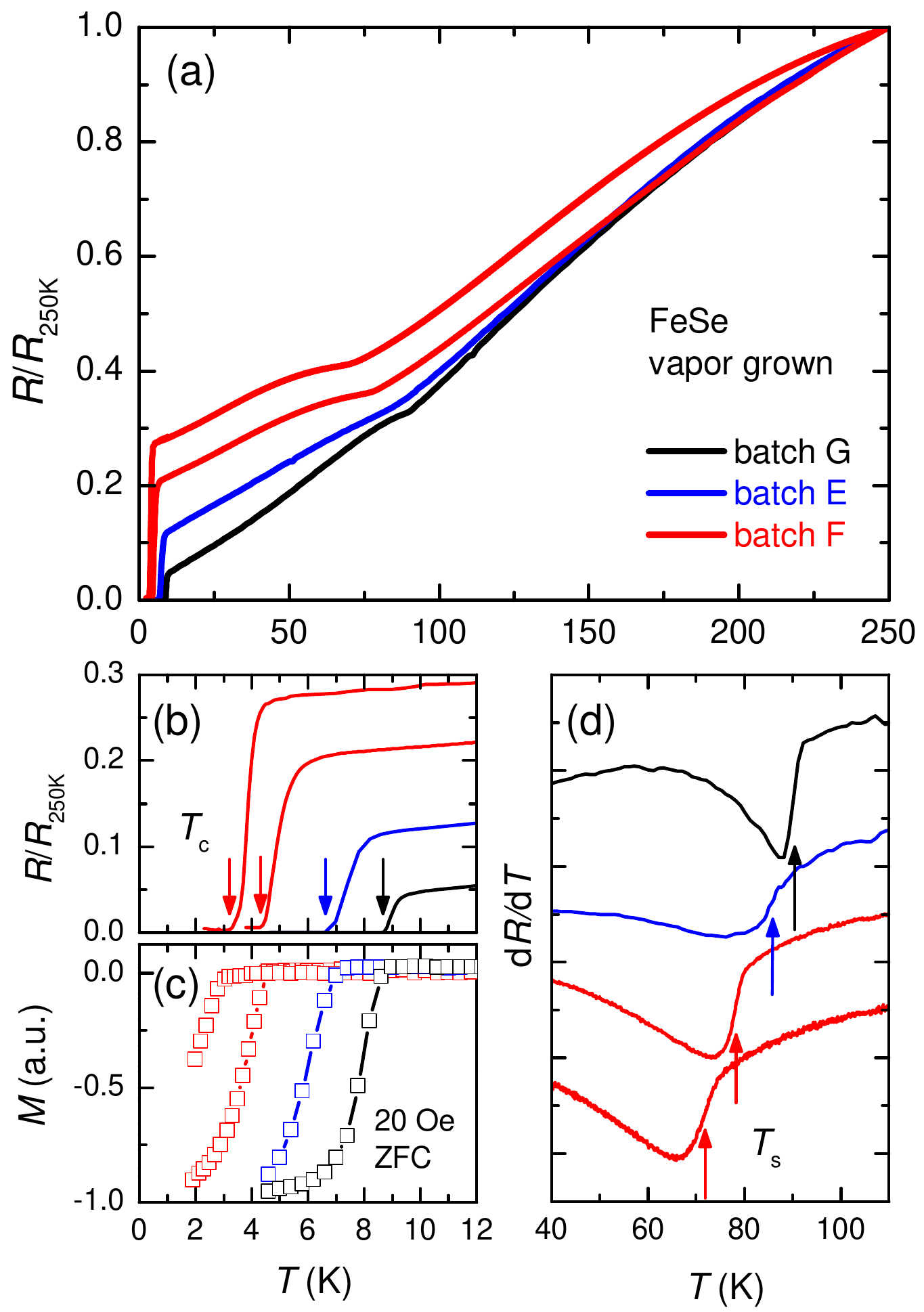}
	\caption{(a) Normalized electrical resistance of four samples of FeSe with large variations in their residual resistivity ratios. (b) Low-temperature resistance data on an expanded scale and (c) zero-field cooled magnetization of the same samples. $T_{\mathrm c}$ as defined by zero resistivity (vertical arrows in (b)) and by the onset of diamagnetic shielding agree well. (d) Temperature derivative of the resistance $dR/dT$ around the structural transition. $T_{\mathrm s}$ is inferred by the mid-point of the step-like anomaly of $dR/dT$ and indicated by arrows. Curves in (d) are offset for clarity.}
	\label{fig:4}
\end{figure}

Fig. 4 shows resistivity curves of four samples, selected for their variation in properties from batches E, F and G. The superconducting transition is more clearly seen in the expanded temperature scale in Fig. 4 (b) and also in the zero-field cooled magnetization in Fig. 4 (c). We define $T_{\mathrm c}$ as the temperature at which the resistivity reaches zero, coinciding with the onset of diamagnetic shielding. The structural transition is clearly seen as a kink in the resistivity data, which results in a step in the derivative $dR/dT$ (Fig. 4 (d)). This feature has frequently been overlooked in data on polycrystalline samples. As evident from the figure, samples with a lower value of their residual resistivity ratio have a lower $T_{\mathrm c}$ and also a lower $T_{\mathrm s}$. Notably, the structural transition remains rather sharp, even when $T_{\mathrm s}$ is decreased by almost 15 K in a sample from batch~F. 

\begin{figure}
	\includegraphics[width=8.6cm]{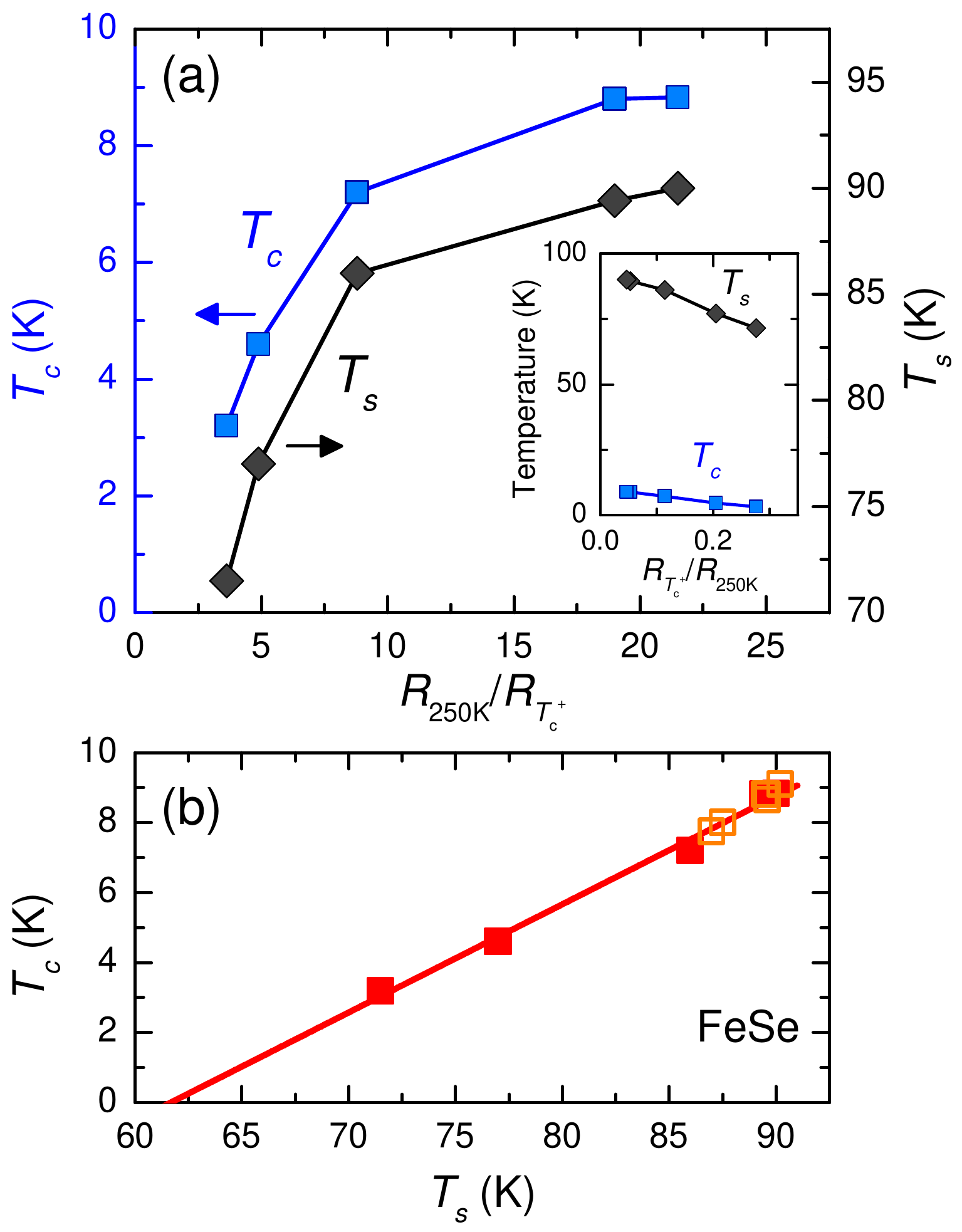}
	\caption{(a) Structural transition temperature $T_{\mathrm s}$ and superconducting transition temperature $T_{\mathrm c}$ as a function of residual resistivity ratio (ratio of resistance at 250 K to resistance just above $T_{\mathrm c}$) for different samples. The inset shows the transition temperature as a function of the inverse residual resistivity ratio. (b) $T_{\mathrm c}$ as a function of $T_{\mathrm s}$ for various samples. Closed symbols show data from panel (a), and open symbols represent data on samples grown as part of earlier studies\cite{Boehmer2013,Boehmerthesis}, for which the transition temperatures were determined using a thermodynamic bulk probe, namely high-resolution thermal expansion. The straight line indicates a linear correlation.}
	\label{fig:5}
\end{figure}

The correlation between the RRR value, $T_{\mathrm s}$ and $T_{\mathrm c}$ is summarized in Fig. 5. Both $T_{\mathrm s}$ and $T_{\mathrm c}$ have very similar dependences on RRR value and both seem to saturate around $\mathrm{RRR}\gtrsim20-25$, suggesting that very pure samples would show $T_{\mathrm s}\sim90$ K and $T_{\mathrm c}\sim9$ K. Remarkably, there is a linear relation between $T_{\mathrm s}$ and $T_{\mathrm c}$ when the sample-to-sample variation is taken as an implicit parameter (Fig. 5 (b)). An extrapolation of this relation indicates that $T_{\mathrm s}=64$ K corresponds to the complete suppression of superconductivity.  Here, we also compare with samples from five different batches associated with Refs. \onlinecite{Boehmer2013,Boehmerthesis}. For these samples, $T_{\mathrm s}$ and $T_{\mathrm c}$ were determined by high-resolution thermal expansion, which is a thermodynamic bulk probe and very reliable and sensitive in the detection of phase transitions. All data fall on the same curve indicating that the relation between $T_{\mathrm s}$ and $T_{\mathrm c}$ is robust.
The inset in Fig. 5 (a) shows the transition temperatures as a function of the inverse, 1/RRR. 1/RRR, as a measure of disorder scattering, could be considered a 'tuning parameter' here, which significantly decreases both $T_{\mathrm s}$ and $T_{\mathrm c}$. A clear question, yet to be resolved, is the microscopic origin of this RRR variation. 

It is interesting to compare the evolution of transition temperatures in the inset of Fig. 5 (a) with the effects of common tuning parameters. For example, chemical substitution with sulfur or tellurium also suppresses $T_{\mathrm s}$, but slightly increases $T_{\mathrm c}$ for low substitution values\cite{Mizuguchi2009}. Similarly, hydrostatic pressure decreases $T_{\mathrm s}$ and increases $T_{\mathrm c}$ initially \cite{Miyoshi2014,Knoener2015,Kaluarachchi2016}. Recently, it was shown that a dose of 2.5 MeV electron irradiation producing approximately $\sim0.1\%$ Frenkel pairs per formula unit, decreases $T_{\mathrm s}$ by 0.9 K and increases $T_{\mathrm c}$ by 0.4 K (Ref. \onlinecite{Teknowijoyo2016}), similarly to the effects of pressure and Te or S substitution, but different from the sample-to-sample variation observed here. This indicates that origin of the variation of RRR, $T_{\mathrm s}$ and $T_{\mathrm c}$ in our samples is more complex than simple vacancy/interstitial-type defects, suggesting instead more extended defects such as dislocation lines. 

Notably, the structural transition in FeSe is almost as sensitive to disorder as superconductivity. This is reminiscent of other Fe-based compounds. For example, $T_{\mathrm s}$ of BaFe$_2$As$_2$ was found to increase from 136 K to 142 K on annealing \cite{Ishida2011}. Such a strong disorder dependence is consistent with an electronically-driven structural transition, often referred to as electronic nematicity in this class of materials. It would be interesting to determine whether the pressure-induced magnetic transition of FeSe\cite{Bendele2010,Bendele2012,Kothapalli2016} is similarly sensitive to the sample-to-sample variation and correlates with $T_{\mathrm s}$ and $T_{\mathrm c}$ under these conditions. Finally, we note that our results stress the importance of carefully separating small changes of $T_{\mathrm c}$ due to changes in preparation conditions from effects of chemical substitution.

\section{Summary}

In summary, we studied the variation of sample morphology, transition temperatures and residual resistivity ratio with small modifications in the preparation conditions of vapor-grown FeSe single crystals. We find that some excess Fe in the starting composition suppresses the formation of the competing hexagonal Fe$_7$Se$_8$ phase, however, its exact amount is less important. The growth seems strongly influenced by the temperature conditions. We find that the highest and most uniform quality crystals are produced with an Fe:Se ratio of 1.1:1 and a small, well-controlled temperature gradient of 350$^\circ$C-390$^\circ$C. Both transition temperatures $T_{\mathrm s}$ and $T_{\mathrm c}$ are found to decrease sensitively with residual resistivity ratio, however, no correlation between $T_{\mathrm c}$ and sample composition was found. In particular, the high sensitivity of $T_{\mathrm s}$ to disorder is consistent with the structural transition being of electronic origin. 

\section*{Acknowledgements}
 We are grateful to S. L. Bud'ko for the critical reading of the manuscript and his valuable comments. This work was carried out at the Iowa State University and supported by the Ames Laboratory, US DOE, under Contract No. DE-AC02-07CH11358. A.E.B. also acknowledges support from the Helmholtz association via PD-226.	

%


\end{document}